\documentstyle[prl,tighten,aps,epsfig,graphics,multicol,amstex]{revtex}

\draft
\begin{document}

\title{Bounds on relative entropy of entanglement for multi-party systems}
\author{M. B. Plenio${}^{1}$ and V. Vedral${}^{1,2}$}
\address{${}^{1}$ Optics Section, The Blackett Laboratory, Imperial College,
London SW7 2BW, UK}
\address{${}^{2}$ Centre for Quantum Computation,
Clarendon Laboratory, Univ. of Oxford, Oxford OX1 3PU,UK}
\maketitle
\begin{abstract}
We present upper and lower bounds to the relative entropy of
entanglement of multi-party systems in terms of the bi-partite
entanglements of formation and distillation and entropies of
various subsystems. We point out implications of our results to
the local reversible convertibility of multi-party pure states and
discuss their physical basis in terms of deleting of information.
\end{abstract}

\draft
\begin{multicols}{2}

\section{Introduction}
The quantification of entanglement is a long standing problem in
quantum information theory \cite{Plenio V 98}. Early work in this
field focused on bi-partite entanglement, but in this paper we
address the problem of quantifying entanglement for multi-party
systems. This is an interesting and complex issue since it is not
always evident how the existing measures for bi-partite systems
can be generalized to the multi-partite case. The three most
promising ideas for quantifying entanglement of multi-party
systems are the entanglement of distillation \cite{Bennett BPS
96}, the entanglement of formation \cite{Bennett BPS 96,Bennett
DSW 96,Wootters 98} and the relative entropy of entanglement
\cite{Vedral PRK 97,Vedral PJK 97,Vedral P 98,Vedral 99,Plenio VP
00}. In the bi-partite setting, the meaning of both entanglement
of formation and the entanglement of distillation is quite clear.
However, these two measures do not have an entirely
straightforward meaning when one considers mulit-partite
entanglement. Let us consider the entanglement of distillation
first. This would be the number of "maximally" entangled states
that the parties could distill asymptotically by local operations
and classical communication (LOCC) from a given multi-party state.
The central issue here is the definition of maximally entangled
states, i.e. to what maximally entangled states should one distill
to. In the multi-partite setting this question has not found an
answer yet, and it is indeed not clear what this answer would be.
In the bi-partite case a maximally entangled state is the one that
allows to obtain any other pure state with certainty by LOCC and,
in fact, it can be shown that the conversion is asymptotically
reversible. For more than two qubits no such single pure state
exists and it is unknown whether there is a finite set of states
with this property \cite{Bennett PRST 99,Linden PSW 99,Wu Z 00}.
For three qubit pure states one may believe that the minimal set
from which every other pure state can be generated reversibly
consists of GHZ and EPR states and it would then be reasonable to
denote this set as the 'maximally entangled set' to which to which
everything else should be distilled. However, it is not known
whether the set of GHZ and EPR states is indeed sufficient to
generate any other pure state asymptotically reversibly and there
are indications that further states have to be added \cite{Galvao
PV 00}. The same issues appear when we talk about the entanglement
of formation. This measure tells us how many maximally entangled
state the parties have to share a priori to be able to
asymptotically create a given mixed state by LOCC.

In general, we would hope that at least for pure states there is a
minimal reversible entanglement generating set (MREGS), to which
all multi-parti states can be distilled and from which all of them
can be reversibly created. As these sets are currently unknown and
may be very complex, we would like to avoid them and instead use
the relative entropy of entanglement. This measure can easily be
generalised to multi-party states. Since for bi-partite systems
the relative entropy of entanglement is an upper bound to
distillable entanglement we would expect this to be the case in
general.

We now proceed to show that this intuition is indeed correct and
derive various upper and lower bounds of the relative entropy of
entanglement for general multi-party states in terms of the
bi-partite entanglements of formation and distillation and
entropies of various subsystems. We discuss the physical meaning
of our bounds in terms of deleting information and present some
relevant applications.

\section{The relative entropy and entanglement: Definitions and useful formulae}

In this paper we will employ a particular measure of entanglement
which is commonly called the relative entropy of entanglement
\cite{Vedral PRK 97,Vedral PJK 97,Vedral P 98,Vedral 99,Plenio VP
00}. This measure can be defined for an arbitrary number of
parties by the following formula
\begin{equation}
    E(\sigma) = min_{\rho\in {\cal D}} S(\sigma||\rho)
    \label{relent}
\end{equation}
where ${\cal D}$ is a set of disentangled (separable) states and
where $S(\sigma||\rho)=tr \{\sigma\log\sigma-\sigma\log\rho\}$ is
the quantum relative entropy \cite{Wehrl 78}. For the purpose of
this paper, we assume that ${\cal D}$ is the set of the states
that can be created locally, i.e. it is {\em fully} separable.
Also, by $E_n(\sigma)$ we will always denote the relative entropy
of entanglement for n-party systems with respect to the set of
fully separable states.


Many of the results in this paper are based on the following
inequality \cite{Plenio VP 00} that holds true for any state
$\sigma_{AB}$ and any separable state $\rho_{AB}$ of two parties.
\begin{equation}
    S(\sigma_{AB}||\rho_{AB}) - S(\sigma_A||\rho_A) \ge S(\sigma_A) -
    S(\sigma_{AB})
    \label{ineqbi}
\end{equation}
This inequality can be generalized directly to multi-partite
entangled systems. For tri-partite systems, for example, we find
that for any state $\sigma_{ABC}$ and any tri-separable state
$\rho_{ABC}$, the following inequality
\begin{equation}
    S(\sigma_{ABC}||\rho_{ABC}) - S(\sigma_{AB}||\rho_{AB}) \ge
    S(\sigma_{AB}) - S(\sigma_{ABC})
    \label{ineqtri}
\end{equation}
is satisfied. We now show how to use this result to derive upper
and lower bounds on the relative entropy of entanglement of
many-party states.

\section{Bounds on the relative entropy of entanglement}

The aim of this section is the derivation of upper and lower
bounds of the relative entropy of entanglement for $n$ parties in
terms of the relative entropies for smaller numbers of subsystems.
We begin by considering pure tri-partite states and later
generalize the results to the case of arbitrarily many subsystems.

{\bf Theorem 1:} For any pure tri-partite states $\sigma_{ABC}$ we
find
\end{multicols}
\begin{eqnarray}
    && \max\{ E_2(\sigma_{AB}) + S(\sigma_{AB}),E_2(\sigma_{AC})
     +  S(\sigma_{AC}),E_2(\sigma_{BC})
    + S(\sigma_{BC})\} \le E_3(\sigma_{ABC}) \label{theorem1} \\
    && E_3(\sigma_{ABC}) \le  \min \{S(\sigma_A) +
    S(\sigma_B), S(\sigma_A)+ S(\sigma_C), S(\sigma_B) + S(\sigma_C) \}
    \label{theorem2}
\end{eqnarray}
\begin{multicols}{2}
{\bf Proof:} We begin with the proof of the first inequality. We
will employ the fact that $E_3$ is an entanglement monotone, i.e.
it does not increase under local operations. Consider now three
parties A, B and C that wish to create an arbitrary pure quantum
state $\sigma_{ABC}$. One possible procedure is that Alice creates
the state locally and then compresses the particles that should go
to Bob (with the efficiency $S(\sigma_B)$) and those that should
go to Charles (with the efficiency $S(\sigma_C)$) and then uses
$S(\sigma_B)$ shared singlets between herself and Bob and
$S(\sigma_C)$ shared singlets between herself and Charles to
teleport these particles to them. Therefore they have consumed
$S(\sigma_C) + S(\sigma_B)$ ebits in total to create this state.
As we know that $E_3$ is an entanglement monotone, we find
\begin{equation}
    E_3(\sigma_{ABC}) \le S(\sigma_B) + S(\sigma_C) \; .
\end{equation}
Permuting the indices cyclically we find the sharpest bound
\begin{eqnarray}
    E_3(\sigma_{ABC}) & \le & \min \{S(\sigma_A) +
    S(\sigma_B), \nonumber\\
    &  & S(\sigma_A)+ S(\sigma_C), S(\sigma_B) + S(\sigma_C) \} .
\end{eqnarray}

Let us now proceed to prove the second inequality. This can easily
be performed using Eq. (\ref{ineqbi}). Given a pure state
$\sigma_{ABC}$, we obtain from this inequality that
\begin{equation}
    S(\sigma_{ABC}||\rho_{ABC}) \ge S(\sigma_{AB}||\rho_{AB}) +
    S(\sigma_{AB}) \; .
    \label{lemma1a}
\end{equation}
If we replace $\rho_{ABC}$ by the closest tri-separable state to
$\sigma_{ABC}$ which we denote by $\rho_{ABC}^{*}$ then we find
\begin{equation}
    E_3(\sigma_{ABC}) = S(\sigma_{ABC}||\rho_{ABC}^*) \ge
    S(\sigma_{AB}||\rho_{AB}^*) + S(\sigma_{AB})
    \label{lemma1b}
\end{equation}
where $\rho_{AB}^* \equiv tr_C\{ \rho_{ABC}^* \}$. As
$\rho_{AB}^*$ is evidently separable we immediately have that
$S(\sigma_{AB}||\rho_{AB}^*)\ge E_2(\sigma_{AB})$ so that
\begin{equation}
    E_3(\sigma_{ABC}) =  S(\sigma_{ABC}||\rho_{ABC}^*) \ge E_2(\sigma_{AB})
    + S(\sigma_{AB}) \label{lemma1c}
\end{equation}
If we permute the indices cyclically we get three inequalities and
obtain the sharpest bound
\begin{eqnarray}
    E_3(\sigma_{ABC} ) & \ge & \max\{E_2(\sigma_{AB})
    + S(\sigma_{AB}),E_2(\sigma_{AC}) \nonumber\\
    & + & S(\sigma_{AC}),E_2(\sigma_{BC})
    + S(\sigma_{BC})\} \; {}_{\Box}.
\end{eqnarray}

{\bf Corollary:} For any pure tri-partite states $\sigma_{ABC}$ we
find
\end{multicols}
\begin{equation}
    \frac{2}{3} \left( S(\sigma_A) + S(\sigma_B) + S(\sigma_C) \right)
    \ge E_3(\sigma_{ABC}) \ge \frac{1}{3} \left( E_2(\sigma_{AB}) +
    E_2(\sigma_{AB}) + E_2(\sigma_{AB} \right) + \frac{1}{3}\left( S(\sigma_A)
    + S(\sigma_B) + S(\sigma_C) \right) \label{corro}
\end{equation}
\begin{multicols}{2}
{\bf Proof:} The Corollary follows directly from the inequalities
stated in the theorem by taking the average over all combinations
of two parties ${}_{\Box}$.

One may wonder for which states the bounds presented in the
theorem and the corollary are saturated. While we do not have a
general answer to this question, it is straightforward to show
that both inequalities \ref{theorem1} and \ref{theorem2} are
saturated for GHZ-like states
($\alpha|000\rangle+\beta|111\rangle$ or local unitary
transformations) and any bi-partite pure state (where the third
subsystem is disentangled from the first two). For the W-state
given by $|W\rangle = (|100\rangle + |010\rangle +
|001\rangle)/\sqrt{3}$ we find that the lower bound \ref{theorem2}
is saturated only as we find $E_3(W)=2\log 3-2$ \cite{Galvao
priv}. In fact for all states of the form $|\psi\rangle =
e|100\rangle + f|010\rangle + f |001\rangle$ \cite{Galvao PV 00} e
find that the upper bound \ref{theorem2} is not saturated .

It would be interesting to know whether one can sharpen the upper
bound presented in the corollary can be sharpened further to
\begin{equation}
    E_3(\sigma_{ABC}) \le \frac{1}{2} \left( S(\sigma_A) +
    S(\sigma_B) + S(\sigma_C) \right)
\end{equation}
We have no counter-example but no proof of this conjecture either.
It is, however, clear that the coefficient in front of the
entropies cannot be sharpened further because this upper bound is
saturated by, for example, EPR states and GHZ states.

The inequalities we have presented here for tri-partite systems
generalize straightforwardly to more than three parties. This is
obtained by first generalizing the inequality \ref{ineqtri} to n
systems and then following the steps leading up to theorems
\ref{theorem1} and \ref{theorem2}.

\section{Reversible entanglement manipulation and the relative entropy of
entanglement}

It is interesting to note that a small modification of the lower
bound presented in Theorem 1 actually becomes an equality for pure
multi-partite states under some further assumptions which we
review now. First of all we will have to replace the relative
entropy of entanglement by its regularized version, ie.
\begin{equation}
    E_3^{\infty} (\sigma) = \lim_{n\rightarrow\infty}
    \frac{E_3(\sigma^{\otimes n})}{n} \; .
\end{equation}
Furthermore let us make the ({\em unproven}) assumption that that
the set of GHZ and the three possible EPR's forms what we called a
MREGS. If this is so, then the following equations must be
satisfied \cite{Linden PSW 99}
\begin{eqnarray}
    S(\sigma_A) &=& g + s_{AB} + s_{AC} \\
    S(\sigma_B) &=& g + s_{AB} + s_{BC} \\
    S(\sigma_C) &=& g + s_{AC} + s_{BC} \\
    E_2^{\infty} (\sigma_{AB}) &=& s_{AB} \\
    E_3(\sigma_{ABC}) &=& g + s_{AB} + s_{AC} + s_{BC}
\end{eqnarray}
where $g$ is the number of GHZs and $s_{AB}$ is the number of
singlets between A and B and so on. For the derivation of the
first four equations see \cite{Linden PSW 99} and \cite{Galvao PV
00} for a slight modification. The last equality follows from the
fact that $E_3^{\infty}$ is an entanglement monotone.  Given these
equations we immediately find that
\begin{eqnarray}
    E_3^{\infty}(\sigma_{ABC}) &=& \frac{1}{3}\left(
    E_2^{\infty}(\sigma_{AB}) +     E_2^{\infty}(\sigma_{AB}) +
    E_2^{\infty}(\sigma_{AB}) \right)    \nonumber \\
    && + \frac{1}{3}\left(
    S(\sigma_{A}) + S(\sigma_{B}) + S(\sigma_{C}) \right)
    \label{lemma1e}
\end{eqnarray}
This equality has an interesting physical interpretation. Namely,
it states that the entanglement of three subsystems is equal to
the entropy due to deleting one of the subsystems plus the
remaining entanglement between the other two subsystems, finally
averaged over all three subsystems. This interpretation of
multi-party entanglement is interesting because it combines the
idea of the persistent entanglement due to loss of classical
information \cite{Eisert 00,Henderson 00} with Landauer's notion
of deleting information and increasing entropy \cite{VPerasure}.
The persistent entanglement tells us how much entanglement is left
when we erase information in a particle and Landauer's erasure
tells us the entropy increase due to this erasure. Their sum gives
the total entanglement.

\section{Discussion and Conclusions}

We have discussed in this paper how to generalize the relative
entropy of entanglement to states involving more than two parties.
Furthermore we have argued that this measure has a very natural
generalization to multi-party states, while the entanglement of
formation and the entanglement of distillation cannot be so easily
generalized. In spite of this advantage of the relative entropy of
entanglement, however, it is still very difficult to compute this
quantity and no `closed formula' exists even for bi-partite
states. It is therefore very important to provide upper and lower
bounds for the relative entropy of entanglement which is the main
result of this paper. We have also discussed the states which
saturate either the upper or the lower bound. Finally, we have
discussed the circumstances under which the lower bound becomes an
equality and related this to Landauer's notion of information
deletion and entropy increase.

\acknowledgements We would like to thank E.F. Galvao and L.
Henderson for useful discussions. We are grateful to the Erwin
Schr{\"o}dinger Institute and Gasthof Hopferl in Vienna for their
hospitality during our visit when this work was performed. We
acknowledge financial support from the Erwin Schr{\"o}dinger
Institute, Elsag-Bailey, EPSRC, Hewlett-Packard, The Leverhulme
Trust, and the EQUIP programme of the European Union.

\end{multicols}

\begin{thebibliography}{99}
\bibitem{Plenio V 98} M.B. Plenio and V. Vedral, Contemp. Phys. \textbf{39},
431 (1998).

\bibitem {Bennett BPS 96} C. H. Bennett, H. J. Bernstein, S. Popescu, B.
Schumacher, Phys. Rev. A \textbf{53}, 2046 (1996).

\bibitem{Bennett DSW 96} C.H. Bennett, D. DiVincenzo, J. Smolin
and W.K. Wootters, Phys. Rev. A {\bf 53}, 3824 (1996).

\bibitem{Wootters 98} W.K. Wootters, Phys. Rev. Lett. {\bf 80},
2245 (1998).

\bibitem{Vedral PRK 97}  V. Vedral, M.B. Plenio, M.A.
Rippin, and P.L. Knight, Phys. Rev. Lett. {\bf 78}, 2275 (1997).

\bibitem{Vedral PJK 97} V. Vedral, M.B. Plenio, K. Jacobs and P.L. Knight,
Phys. Rev. A {\bf 58}, 4452 (1997).

\bibitem {Vedral P 98} V. Vedral and M. B. Plenio, Phys. Rev. A \textbf{57}, 1619 (1998).

\bibitem {Vedral 99} V. Vedral, Phys. Lett. A \textbf{262}, 121 (1999).

\bibitem{Plenio VP 00} M.B. Plenio, S. Virmani and P. Papadopoulos,
J. Phys. A {\bf 33}, L193 (2000).

\bibitem {Bennett PRST 99} C. H. Bennett, S. Popescu, D. Rohrlich, J. A.
Smolin, A. V. Thapliyal, Los Alamos e-print quant-ph/9908073.

\bibitem {Linden PSW 99} N. Linden, S. Popescu, B. Schumacher, M. Westmoreland,
Los Alamos e-print quant-ph/9912039.

\bibitem {Wu Z 00} S. Wu and Y. Zhang, Los Alamos e-print quant-ph/0004020.

\bibitem{Galvao PV 00} E.F. Galvao, M.B. Plenio and S. Virmani,
quant-ph/0008089, accepted for publication in J. Phys. A.

\bibitem{Galvao priv} E.F. Galvao and L. Henderson, private communication of a numerical result.

\bibitem {Wehrl 78} A. Wehrl, Rev. Mod. Phys. \textbf{50}, 221 (1978).

\bibitem {Eisert 00} J. Eisert, T. Felbinger, P. Papadopoulos, M.B. Plenio, and M. Wilkens.
Phys. Rev. Lett. {\bf 84}, 1611 (2000).

\bibitem{Henderson 00} L. Henderson and V. Vedral, Phys. Rev.
Lett. {\bf 84}, 2263 (2000).

\bibitem{VPerasure} V. Vedral, Proc. Roy. Soc. London A {\bf 456}, 969 (2000); M. B.
Plenio, Phys. Lett. A {\bf 263}, 281 (1999).

\end{thebibliography}
\end{document}